\documentclass[aps,prl,showpacs,groupedaddress,twocolumn]{revtex4}
\usepackage{bm}
\usepackage{graphicx}

\newcommand{\dt}{\textrm{d}t}

\newcommand{\x}{\bm{x}}
\renewcommand{\j}{\bm{j}}
\newcommand{\J}{\bm{J}}

\renewcommand{\S}{\bar{S}}
\newcommand{\n}{\bar{n}}
\newcommand{\Tr}{\mbox{Tr}}
\newcommand{\jj}{\bm{\bar{\jmath}}}
\newcommand{\bes}{\begin{eqnarray}}
\newcommand{\ees}{\end{eqnarray}}
\newcommand{\be}{\begin{equation}}
\newcommand{\ee}{\end{equation}}
\begin{document}
\title{Current in open quantum systems}
\author{Ralph Gebauer}
\affiliation{The Abdus Salam International Centre for Theoretical
Physics (ICTP), 34014 Trieste, Italy\\
INFM Democritos National Simulation Center, 34013 Trieste, Italy} 
\author{Roberto Car}
\affiliation{Department of Chemistry and Princeton Materials
Institute, Princeton University,  Princeton, New Jersey 08540 } 
\date{\today} 
\begin{abstract}
We show that a dissipative current component is present 
in the dynamics generated by a Liouville-master equation, 
in addition to the usual component 
associated with Hamiltonian evolution.   
The dissipative component originates from coarse graining in time, 
implicit in a master equation, and needs to be
included to preserve current continuity. 
We derive an explicit expression for the
dissipative current in the context of the Markov approximation. 
Finally, we illustrate our approach with a
simple numerical example, in which a quantum particle is coupled
to a harmonic phonon bath and dissipation is described by the Pauli master
equation.    
\end{abstract}
\maketitle

Open quantum systems are encountered in many domains of physics, ranging from 
condensed matter physics \cite{Feynman, Caldeira} to quantum optics
\cite{Haken}, chemical physics \cite{vanKampen}
and  NMR theory \cite{Redfield}. In all 
these fields one is interested in the reduced dynamics of a quantum system in contact 
with its environment. The reduced dynamics is usually approximated by a master 
equation. A simple example is the Pauli master equation (PME) \cite{Pauli} in which the diagonal 
elements $\S_{ll}(t) = \left< l \left| \S(t) \right| l \right>$  
of a coarse grained reduced density operator   
$\S(t)$ in the basis of the system eigenstates evolve in time according to: 
\be
\label{PME}
\frac{d \S_{ll}}{dt} = \sum_{m \ne l} \left( W_{l m} \S_{mm} - W_{m l} \S_{ll} \right).  
\ee

The transition probabilities $W_{ij}$ are given by 
Fermi's golden rule expression in terms  of the coupling to the
bath. The  $W_{ij}$ satisfy 
detailed balance, which guarantees that the 
thermal equilibrium distribution is the stationary solution of
Eq.~(\ref{PME}). A difficulty with  
Eq.~(\ref{PME}) is that it describes a charge (or probability) density
evolution that does not seem to  
satisfy the current continuity equation, given that   
the current associated with eigenstates is divergenceless. This
apparent violation of current  
continuity makes the use of Eq.~(\ref{PME}) problematic to model
electron transport kinetics \cite{Frensley, Fischetti}, although it
has been argued that under  steady state conditions 
Eq.~(\ref{PME}) can still be used to simulate electron transport in
very small devices \cite{Fischetti}. 

Eq.~(\ref{PME}) is a special case of a Liouville master equation of
the form: 
\be
\label{ME}
\frac{d \S(t)}{dt} = -i \left[ H, \S(t)\right]
+ {\cal C} [\S(t)],
\ee
where $H$  is the Hamiltonian of the system and  ${\cal
  C}[\S(t)]$  describes the dissipative part of 
the dynamics after eliminating bath memory effects by coarse-graining in time (Markov
approximation) \cite{Louisell, Cohen-Tannoudji}. 
Eq.~(\ref{ME}) is derived by assuming a harmonic bath and by treating the system-bath interaction
to 2nd order of perturbation theory. The limits of validity of such approximation
and of the Markov approximation requiring separation of the time scales of system and bath, 
have been amply discussed in the literature \cite{Cohen-Tannoudji, Bloch}.  

Eq.~(\ref{ME}) applies to a generic interacting $N-$particle quantum system described by a
$N-$particle coarse-grained reduced density operator $\S$. 

In what follows we adopt atomic units ($e=m=\hbar=1$). We indicate coarse-grained
quantities by a bar. The coarse-grained charge
density 
$\n(\x;t)$ is given  by $\n(\x;t) = \mbox{Tr} \left\{ \S(t)
\hat{\rho}(\x)\right\}$ in terms of the charge density operator
$\hat{\rho}(\x) = \sum_{l=1,N} \delta(\x - \hat{\x}_l)$. The coarse-grained Hamiltonian
current density is given by
$\jj_H (\x;t) = \mbox{Tr} \left\{ \S(t) \hat{\J}(\x)
\right\}$, in terms of the current density operator $\hat{\J}(\x) =
\frac1{2} \sum_{l=1,N} \left[ \hat{\bm{p}}_l \delta(\x - \hat{\x}_l) +
  \delta(\x - \hat{\x}_l) \hat{\bm{p}}_l \right]$. If the
potential in $H$ is local in space, then the
continuity equation $\dot{\n}(\x;t) = - \nabla \cdot \jj_H(\x;t)$ is 
evidently satisfied if \cite{Frensley}:
\be
\label{cond}
\mbox{Tr}\left\{ {\cal C}[\S(t)] \hat{\rho}(\x) \right\} = 0.
\ee  

Generally master equations in which the dissipative dynamics leads 
asymptotically to the thermal equilibrium distribution do not satisfy
Eq.~(\ref{cond}) \footnote{A remarkable exception being the Boltzmann
transport  equation, where the scattering occurs between planewaves.}. This is  
at variance with the exact reduced dynamics which satisfies the continuity equation if 
the coupling to the bath is local. 

In this letter we show that assertions on the violation of continuity 
by master equations that do not satisfy Eq.~(\ref{cond}) are indeed incorrect, and that   
the coarse-grained (Markovian) dynamics exactly satisfies current continuity, as 
the true reduced dynamics does if the coupling to the bath is
{\em local}. This is because the following equation holds:
\be
\label{cond1}
\mbox{Tr}\left\{ {\cal C}[\S(t)] \hat{\rho}(\x) \right\} = \dot{\bar{n}}_{\cal C}(\x) = - \nabla
\cdot \jj_{\cal C} (\x;t)
\ee
In Eq.~(\ref{cond1}) $\jj_{\cal C}$ is a {\em
dissipative current} originating from the action of the bath on the system
on the {\em coarse grained} time scale of Markovian dynamics.
This current contribution, which has been ignored so far, is of
the same order of the Hamiltonian current in strongly inhomogeneous systems.  
We also provide an expression for $\jj_{\cal C}$, whose calculation 
only requires knowledge of the coarse-grained density operator $\S(t)$, 
of the eigenvalues and eigenvectors of the system Hamiltonian $H$, of
the Hamiltonian $R$ of the harmonic reservoir, and of the coupling $K$ between the system
and the bath.   

The derivation of a master equation of the form (\ref{ME}) 
can be found in textbooks \cite{Louisell}. Here we limit ourselves to 
pointing out the key elements of this derivation
that are important for our discussion of current
continuity.
 
Let S(t) be the reduced density operator in absence of coarse graining. 
As usual, to separate the effect of the bath from system dynamics, it is convenient to 
work in the interaction picture (IP): $S(t) =
e^{-iH(t-t_0)}s(t)e^{iH(t-t_0)}$, where
$t_0$ is a time 
preceding $t$  and $s(t)$  is the reduced density operator in the IP. Taking the time derivative 
of $S(t)$, the reduced density operator in the Schr\"odinger picture (SP), one gets:
\be
\label{TD}
\frac{d S(t)}{dt} = -i\left[H,S(t)\right] + e^{-iH(t-t_0)}
\frac{d s(t)}{dt} e^{iH(t-t_0)}.
\ee
Eq.~(\ref{TD}) governs the exact dynamics of the system. 
The first term on the right hand side (r.h.s.)  corresponds to the usual
Hamiltonian dynamics, which satisfies current continuity if the
potentials in $H$ are local. For the second term on the r.h.s.~one can
show that  $\mbox{Tr}\left\{ e^{-iH(t-t_0)} \frac{d s(t)}{dt}
e^{iH(t-t_0)} \hat{\rho}(\x) \right\}=0$, if the interaction potential
with the bath is local. Therefore, this term does not modify the
instantaneous distribution of charges, and the exact dynamics satisfies
current continuity. 

In order to convert Eq.~(\ref{TD}) into a master equation of the form
(\ref{ME}), one replaces the time-derivative $\frac{d s(t)}{dt}$ in
(\ref{TD}) by a finite difference approximation $\frac{\Delta
  s}{\Delta t} = \frac{s(t) - s(t_0)}{t - t_0}$ calculated to leading
order of perturbation theory using Fermi's golden rule, and then one takes the limit $\Delta t
\rightarrow 0$ (corresponding to {\em coarse-graining} in time) in
the equation for system dynamics, {\em i.e.} one writes for the
coarse-grained density operator $\bar{S}(t)$:
\bes
\label{limit}
\frac{d \S(t)}{dt} &=& -i \left[ H, \S(t)\right] + \lim_{t_0
  \rightarrow t} e^{-i H (t-t_0)} \frac{\Delta s}{\Delta t} e^{i H
  (t-t_0)} \nonumber \\
&=& -i \left[ H, \S(t)\right] + \lim_{t_0
  \rightarrow t} \frac{\Delta s}{\Delta t}
\ees

We recall that Fermi's golden rule expression for $\frac{\Delta
s}{\Delta t}$ is independent of $\Delta t$. The limit for $\Delta t
\rightarrow 0$ is possible because $\Delta t$, although large
relative to inverse system frequencies for Fermi's golden rule to be
applicable, is {\em small} compared to the relaxation time. Thereafter all
reference to $t_0$ disappears from Eq.~(\ref{limit}) and the future
evolution of the system is entirely determined by its present and not
by its past (Markov approximation). Eq.~(\ref{limit}) is the same as
Eq.~(\ref{ME}) \cite{Cohen-Tannoudji,Louisell}.

To prove that Eq.~(\ref{ME}) satisfies charge continuity even when Eq.~(\ref{cond})
is not satisfied, we compute the time
derivative of the coarse-grained charge density:
\bes
\label{ndot}
\frac{d \n (\x)}{dt} &=& \Tr \left\{ \frac{d \S(t)}{dt} \hat{\rho}(\x)
\right\}  \\
&=& -i \Tr \left\{ \left[ H,\S(t) \right] \hat{\rho}(\x)\right\} + 
\Tr \left\{ \lim_{\Delta t \rightarrow 0} \frac{\Delta s}{\Delta t} 
\hat{\rho}(\x) \right\}.\nonumber
\ees

The first term on the r.h.s.~of Eq.~(\ref{ndot}) gives: $-i
\Tr \left\{ \left[ H,\S(t)\right] \hat{\rho}(\x)\right\} = - \nabla
\cdot \jj_H(x;t)$. To compute the remaining term we consider the exact
density operator $S(t') = e^{-i H (t'-t_0)} s(t') e^{i H (t'-t_0)}$, whose
dynamics includes the effect of the bath, and the density operator
$S_H (t')$, whose dynamics is governed by $H$ alone and does not include the
effect of the bath: $S_H(t') = e^{-i H (t'-t_0)} s(t) e^{i H
  (t'-t_0)}$. Here we have assumed that at time $t$ $S(t) = S_H(t)$. Then at time
$t_0$, $S(t_0)=s(t_0)$ and $S_H(t_0)=s(t)$. Thus:
\bes
\Delta s &=& s(t)-s(t_0) = S(t)-S(t_0)-S_H(t)+S_H(t_0) \nonumber \\
&=& \int_{t_0}^t dt' \, \left( \dot{S}(t') - \dot{S}_H(t') \right).
\ees
Since both the exact and the $H-$dynamics satisfy continuity, we can
write
\bes
\label{proof}
\Tr\left\{\Delta s \hat{\rho}(\x)\right\} &=& \int_{t_0}^t dt' \,
\left( \dot{n}(\x;t')-\dot{n}_H(\x;t')\right) \nonumber \\
&=& - \nabla \cdot \int_{t_0}^t dt' \, \left(\j(\x;t')-\j_H(\x;t')\right)
\nonumber \\
&=& - \nabla \cdot \int_{t_0}^t dt' \, \j_{\cal C}(\x;t'),
\ees
where $\j_H = \Tr\left\{ S_H \hat{\J}\right\}$ and $\j=\Tr\left\{S
\hat{\J} \right\}$. In Eq.~(\ref{proof}) the difference between the
exact current, which includes the effect of the bath, and the
$H-$current defines the current $\j_{\cal C} \equiv \j - \j_H$. Using
(\ref{proof}) we obtain the second term on the r.h.s.~of
  Eq.~(\ref{ndot}):
\bes
\label{coarsej}
\lim_{\Delta t \rightarrow 0} &\Tr& \left\{ \frac{\Delta s}{\Delta t}
\hat{\rho}(\x)\right\} = - \lim_{t_0 \rightarrow t} \frac1{\Delta t} \nabla \cdot 
\int_{t_0}^t dt' \, \j_{\cal C}(\x;t) \nonumber \\
&=& - \nabla \cdot \left( \lim_{t_0 \rightarrow t} \frac1{\Delta t}
\int_{t_0}^t dt' \, \j_{\cal C}(\x;t)\right). 
\ees
We identify the term in parenthesis in Eq.~(\ref{coarsej}) with the
coarse-grained current $\jj_{\cal C}$.
Thus we obtain
\be
\label{curr}
\frac{d \n(\x)}{dt} = - \nabla \cdot \jj_H(\x;t) - \nabla \cdot
\jj_{\cal C}(\x;t) = - \nabla \cdot \jj(x;t).
\ee

Eq.~(\ref{curr}) is the main result of this Letter: it shows that the
master equation {\em does not} violate current continuity and that the
coarse-grained physical current $\jj$, {\em i.e.} the current measured
in experiments, is the sum of two contributions, a current $\jj_H$ ,
associated with the charge flow from Hamiltonian propagation
in the Liouville master equation (\ref{ME}), and a current $\jj_{\cal
  C}$, associated with the charge flow due to inelastic collisions
with the bath in the same equation. In the following we show that the
limit in Eq.~(\ref{coarsej}) can be calculated to leading order of
perturbation theory, {\em i.e.} at the same level of approximation
used in the derivation of the master equation itself. The resulting
expression for $\jj_{\cal C}$ is independent of $t_0$. Thus $\jj
(x;t)$ can be calculated from the knowledge of $\S(t)$ at time $t$ only.

For the derivation of an explicit expression for $\jj_{\cal C}$, we
consider a coupling $K$   
of the form $K = \sum_\alpha V^\alpha F^\alpha$,
where $V^\alpha = \sum_{l=1,N} V^{\alpha}_l (\hat{\x}_l)$ is a {\em
local} potential acting on the system, and $F^\alpha$ acts on 
bath variables only. 
The sum is over all the bath eigenmodes (phonons), 
but to simplify the notation we drop the index $\alpha$  in what follows. A standard procedure 
\cite{Louisell} gives $\Delta s$ to second order of perturbation theory:
\bes
\label{DS}
\Delta s = s(t)-s(t_0) &=& -\sum_{mnop}\left\{A_{mn,op} w^-_{op} -
 B_{mn,op} w^+_{mn} \right\} \nonumber \\
&&\times \int_0^{t-t_0} \mbox{d}\xi e^{i(e_m-e_n+e_o-e_p)\xi},
\ees
where the sums are over eigenstates $\left|i\right>$  of $H$  having
 energy $e_i$. The operators
$A$   and $B$  are:
\bes
A_{mn,op}&=&\left|m\right>V_{mn}\left<n\left|o\right>\right.V_{op}\left<p\right|
 s(t_0) \nonumber \\
&& -  \left|o\right>V_{op}\left<p\right| s(t_0)
 \left|m\right>V_{mn}\left<n\right|, \nonumber \\
B_{mn,op}&=& \left|o\right>V_{op}\left<p\right| s(t_0)
 \left|m\right>V_{mn}\left<n\right| \nonumber \\
&& - s(t_0)\left|m\right>V_{mn}\left<n\left|o\right>\right.V_{op}\left<p\right|,
\ees
where $V_{mn} = \left<m\left|V\right|n\right>$, and
\bes
\label{BSD}
w^+_{mn} &=& \int_0^{\infty} e^{i(e_m-e_n)t'} \left<F(t')F(0)\right>_R
 \dt', \nonumber \\
w^-_{op} &=& \int_0^{\infty} e^{i(e_o-e_p)t'} \left<F(0)F(t')\right>_R
 \dt'
\ees
are bath spectral densities, which, for a harmonic bath, can be calculated 
analytically and are zero when $m=n$  ($o=p$). In Eq.~(\ref{BSD}) 
$F(t')$ is in the IP and $\left<\cdots\right>_R$  denotes an average
 over the bath degrees of freedom. 
If $t-t_0$  is large compared to the inverse level spacing $1/(e_i-e_j)$,  the time 
integral in Eq.~(\ref{DS}) leads to selection rules, where the only terms that survive are those with 
$m=p$  and $n=o$ . The resulting $\Delta s$  varies linearly with
 $\Delta t$ . Inserting $\frac{\Delta s}{\Delta t}$ in
 Eq.~(\ref{limit}), one obtains the usual expression for the
 dissipative master equation \cite{Louisell}. 

In order to compute the dissipative current, one has to 
evaluate
\be
\label{derc}
\jj_{\cal C}(\x,t) = \lim_{t_0 \rightarrow t} \frac1{\Delta t}
\int_{t_0}^t dt' \, \Tr \left\{\left( S(t')-S_H(t') \right) \J(\x)
\right\}.
\ee
Using $\left( S(t')-S_H(t') \right) = e^{-i
H (t'-t_0)} \left( s(t')-s(t_0)-(s(t)-s(t_0)) \right) e^{i H
(t'-t_0)}$, one can insert the expression (\ref{DS}) for
$s(t')-s(t_0)$ and $s(t)-s(t_0)$, respectively, into Eq.~(\ref{derc}).

The additional time-integral and the exponentials lead to different
selection rules than in case of Eq.~(\ref{DS}).
Identifying the leading terms,  and 
finally letting $t_0 \rightarrow t$ , we obtain an expression for the
dissipative current: 

\be
\label{DC}
\jj_{\cal C}(\x;t) = i \sum_{k\ne l}\frac1{e_l-e_k} \left({\cal T}_{kl}
-{\cal R}_{kl} \right) \hat{\J}(\x)_{lk},
\ee
where $\hat{\J}(\x)_{lk} \equiv
\left<l\left|\hat{\J}(\x)\right|k\right>$ and ${\cal
  R}_{kl}$,${\cal T}_{kl}$ are given by: 
\bes
{\cal R}_{kl} &=& -\sum_{n \ne k} \left|V_{nk}\right|^2 \bar{S}_{kl}(t)
  w^-_{nk} \nonumber \\
&&- \sum_{n \ne l} \bar{S}_{kl}(t) \left|V_{ln}\right|^2 w^+_{ln},\label{RKL}\\
{\cal T}_{kl} &=& -\sum_{n \ne k} V_{kn} V_{nl} \bar{S}_{ll}(t)
  w^-_{nl} \nonumber \\
&&+\sum_{n \ne k} V_{kn} \bar{S}_{nn} V_{nl} (w^+_{nl} + w^-_{kn}) \nonumber \\
&&- \sum_{n \ne l} \bar{S}_{kk}(t) V_{kn} V_{nl} w^+_{kn}, \label{TKL}\ees

Eq.~(\ref{DC}) is the second important result of this paper and completes the proof that 
$\jj_{\cal C} (\x;t)$ can be calculated from knowledge of $\S(t)$ and microscopic Hamiltonian only. 
Eq.~(\ref{DC}) gives (to 2nd order of perturbation theory) the coarse-grained current 
associated to phonon induced inelastic transitions between 
all pairs of system states. The transition matrix elements depend on (amplitude and phase of)
the coupling potential and on the fluctuations of the bath (via the bath spectral densities). 
The derivation of Eq.~(\ref{DC}) does not require additional assumptions
besides those made in the derivation of the Liouville master equation. In this context  
Eq.~(\ref{DC}) is {\em exact}. Our approach requires an explicit
phonon (bath) model, however simplified.  

Sometimes master equations are used without reference to an explicit
bath model. This is the case, for instance, with the common relaxation time
approximation. This approximation does not violate charge continuity when it is
applied to spatially homogeneous systems. In inhomogeneous systems, however,  
the relaxation time approximation leads to violation of charge continuity
whenever the bath originates a spatially non-uniform instantaneous
change of the coarse-grained charge density, {\em i.e.} whenever 
$\dot{\bar{n}}_{\cal C} (\x) \neq 0$ \cite{Mizuta}.

Finally, we use a simple numerical example to illustrate our theory.
In this example a single quantum particle (electron) is in a one dimensional
potential $U(x)$. The electron is initially placed in the first excited eigenstate of the Hamiltonian 
$H = -\frac1{2} \frac{d^2}{dx^2} + U(x)$. In absence of dissipation the electron would stay in the 
excited state, but due to coupling with a phonon bath it relaxes toward thermal 
equilibrium. In this case, neglecting the imaginary part of the bath spectral densities,
Eq.~(\ref{ME}) reduces to Eq.~(\ref{PME}),
{\em i.e.} the PME, and the density matrix stays  
diagonal while the system approaches equilibrium. Even though the
density matrix does  not carry any current, Eq.~(\ref{DC}) gives a
non-zero current due to the ${\cal T}_{kl}$ terms during  relaxation
dynamics. 

\begin{figure}
\includegraphics[angle=-90,width=\columnwidth]{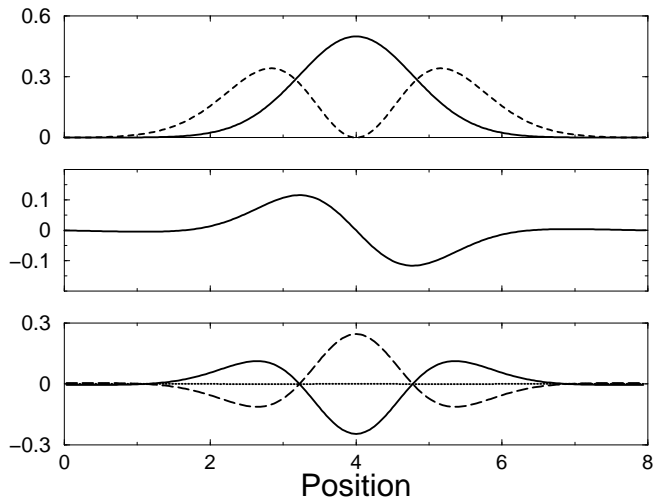}
\caption{\label{NewFig} Upper Panel: Calculated charge densities of
  the ground state (solid line) and first excited state (dashed line). 
  Middle Panel: Dissipative current $\bar{\jmath}_{\cal C} (x)$ at a
  given instant of the propagation, calculated from Eq.~(\ref{DC}). 
  Lower panel: Divergence of $\bar{\jmath}_{\cal C} (x)$ (solid line),
  time derivative of the charge density $\dot{\bar{n}}(x;t)$ at
  the same time (dashed line), and the sum of the two quantities
  (dotted line). It is evident that continuity equation is
  satisfied during relaxation.}
\end{figure}

We take 
\be 
w^+_{nm} = w^-_{mn} = \left\{
\begin{array}{ll}
\bar{n}(e_n-e_m) + 1 & e_n > e_m\\
\bar{n}(e_m-e_n)     & e_n < e_m\\
\end{array} \right. ,
\ee
where $\bar{n}(\omega) = 1/(e^{\frac{\omega}{kT}} - 1)$ is the mean
occupation number of phonons with energy $\omega$.
This simple choice guarantees 
detailed balance. We  assume that $V \equiv \frac{d U(x)}{d x}$  is
the potential that characterizes the coupling with the bath.  
The transition probabilities in Eq.~(\ref{PME}) are given by $W_{lm} = 
2 \left| V_{lm} \right|^2 w^+_{ml}$. 
We represent the space coordinate $x$ on a fine
numerical grid with 256  points.   
In Fig.~\ref{NewFig} we report the calculated charge density $n_0(x)$
of the ground state, the calculated charge density $n_1(x)$  of the
first excited  state, and the calculated time derivative of the charge
density $\dot{\bar{n}}(x;t)$  at a time $t$  during relaxation  
dynamics (in this example we take $kT = 0.4 \left( e_1 - e_0
\right)$). It is clear that there can be no current associated 
to the eigenstates to compensate the change of charge density
$\dot{\bar{n}}(x;t)$.  

In the same figure, we report the dissipative current density $\bar{\jmath}_{\cal
C}(x;t)$ at the same time $t$. It is obtained from
Eq.~(\ref{DC}), where ${\cal R}_{kl} = 0$ because $\bar{S}(t)$ is diagonal.
As we can see, the continuity equation   
$\dot{\bar{n}}(x;t) + \frac{d}{dx} \bar{\jmath}_{\cal C}(x;t) = 0$ is exactly 
satisfied within the numerical accuracy of the calculation.

Numerical integration of Eq.~(\ref{ME}) for a single quantum particle
is feasible, but is not possible in general for interacting many-body
quantum systems. In practical applications to a many-body
quantum system, the latter is usually approximated by a system of
non-interacting particles in an effective single-particle 
Hamiltonian like {\em e.g.} the time dependent Hartree Hamiltonian
\cite{TDH}.    
Then Eq.~(\ref{ME}), the $N-$particle Liouville master equation, can
be reduced to an equation for an effective {\em single-particle}
density operator in which the master term depends quadratically rather
than linearly on the (single-particle) density
operator\cite{Hubner}.  The basic assumptions on perturbative
coupling to the bath and coarse-graining in time that led to
Eq.~(\ref{ME}) remain valid, however, and so is our analysis that led
to Eq.~(\ref{DC}). Also in this case, a proper expression for 
$\jj_{\cal C}(\x;t)$ can be
found, which restores exactly current continuity \cite{MPexpression},
removing a major limitation to the application  of quantum kinetic
approaches to the study of electron transport problems.

\begin{acknowledgments}  
We thank Morrel Cohen, Erio Tosatti, Kieron Burke, and
David Vanderbilt for many stimulating discussions. This work has
been partially supported by DOE under grant DE-FG02-01ER45928 and by ONR under  
grant N00014-01-1-1061.
\end{acknowledgments}


\begin{thebibliography}{99}
\bibitem{Feynman} R.P.~Feynman and F.L.~Vernon, Ann.~Phys. {\bf 24},
  118 (1963).
\bibitem{Caldeira} A.O.~Caldeira and A.J.~Leggett, Physica A {\bf
  121}, 587 (1983).
\bibitem{Haken} H.~Haken, {\em Laser Theory} (Springer, Berlin, 1970).
\bibitem{vanKampen} N.G.~van Kampen, {\em Stochastic Processes in
  Physics and Chemistry} (North-Holland, Amsterdam, 1992).
\bibitem{Redfield} A.G.~Redfield, IBM J.~Res.~Dev. {\bf 1}, 19 (1957).
\bibitem{Pauli} W.~Pauli, {\em Festschrift zum 60.~Geburtstage
  A.~Sommerfeld} (Hirzel, Leipzig, 1928), p. 30.
\bibitem{Frensley} W.R.~Frensley, Rev.~Mod.~Phys. {\bf 62}, 745
  (1990).
\bibitem{Fischetti} M.V.~Fischetti, J.~Appl.~Phys. {\bf 83}, 270
  (1998).
\bibitem{Cohen-Tannoudji}  C.~Cohen-Tannoudji, J.~Dupont-Roc, and
  G.~Grynberg, {\em Atom-photon interactions: basic processes and
  applications}, Wiley, New York (1992). 
\bibitem{Louisell} W.H.~Louisell, {\em Quantum Statistical Properties
  of Radiation} (Wiley, New York, 1973).
\bibitem{Mizuta} H.~Mizuta and G.J.~Goodings,
  J.~Phys.: Cond.~Mat. {\bf 3}, 3739 (1991).  
\bibitem{quantumKinetic} R.~Gebauer and R.~Car, submitted.
\bibitem{Bloch} R.K.~Wangness and F.~Bloch, Phys.~Rev. {\bf 89}, 728
  (1953). 
\bibitem{Fischetti1} M.V.~Fischetti, Phys.~Rev.~B {\bf 59}, 4901
  (1999).
\bibitem{TDH} P.A.M.~Dirac, Proc.~Cambr.~Philos.~Soc. {\bf 26}, 376
  (1930), A.D.~McLachlan, Mol.~Phys. {\bf 8}, 39 (1964).
\bibitem{Hubner} R.~H\"ubner, R.~Graham, Phys.~Rev.~B {\bf 53}, 4870 (1996).
\bibitem{MPexpression} R.~Gebauer and R.~Car, in preparation.
\end{thebibliography}
\end{document}